\documentstyle[11pt,newpasp,twoside,epsf]{article}
\def\edcomment#1{\iffalse\marginpar{\raggedright\sl#1\/}\else\relax\fi}
\reversemarginpar

\begin{document}
\title{Quasar clustering redshift distortion constraints on dark energy}
 \author{Maur\'\i cio O. Calv\~ao, Jo\~ao R. T. de Mello Neto, and Ioav Waga}
\affil{Universidade Federal do Rio de Janeiro, Instituto de F\'\i
sica, C. P. 68528, CEP 21945-970, Rio de Janeiro, RJ, Brazil}

\begin{abstract}
Redshift distortions, both geometrical and kinematical, of quasar
clustering are simulated, for the Two-Degree Field QSO Redshift
Survey (2QZ), showing that they are very effective to constrain
the cosmological density and equation of state parameters,
$\Omega_{m0}$, $\Omega_{x0}$ and $w$. Particularly, it emerges
that, for the cosmological constant case, the test is especially
sensitive to the difference $\Omega_{m0}-\Omega_{\Lambda 0}$,
whereas, for the spatially flat case, it is quite competitive with
future supernova and galaxy count tests, besides being
complimentary to them.
\end{abstract}

\section{Introduction}

Following Alcock \& Paczy\'nski's (1979) and Phillipps' (1994)
lead, we extended Popowski et al. (1998) investigation as follows:
(1) We performed Monte Carlo simulations to obtain the probability
density function and corresponding confidence contours in the
parametric plane $(\Omega_{m0},\Omega_{\Lambda 0})$, comparing
them to other tests; (2) We included a general dark energy
component with constant equation of state parameter $w$,
obtaining, for flat models, the confidence contours in the
$(\Omega_{m0},w)$ plane; (3) We explicitly took into account the
effect of large-scale coherent peculiar velocities (Hamilton 1992;
Matsubara \& Suto 1996). Our calculations are based on the
measured 2QZ distribution function and we consider best fit values
for the amplitude and exponent of the correlation function as
obtained by Croom et al. (2001).

\section{Results}

Our results are summarized in Fig. 1, where the scattered points
represent the maximum likelihood estimates from our simulations
and the solid black dots the ``true'' starting models. The left
panel displays typical $1\sigma$ confidence contours for a SNAP
simulation (dashed line; Goliath et al.\ 2001) and for the
Supernova Cosmology Project (dotted line; Perlmutter et al.\
1999), besides our corresponding contour (solid line), all for
typical cosmological constant ($w=-1$) models. The right panel
displays two sets of confidence contours, all for typical flat
models: the upper ones correspond to $95\%$ level for our
simulations (solid line) and for DEEP (dotted; Newman \& Davis
2000); the lower ones correspond to $1\sigma$ level for our
simulations (solid lines) and for SNAP (dashed)

\begin{figure}
\plotone{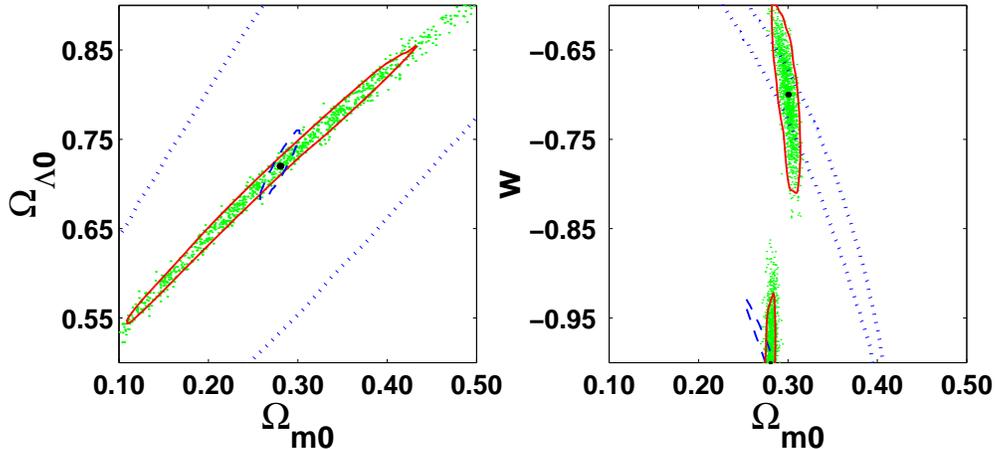} \caption{Constraints for $w=-1$ ($\Lambda$)
models [left panel] and $k=0$ (flat) models [right panel].}
\end{figure}

\section{Discussion and conclusion}

Fig.~1 shows that the test is quite sensitive (when $w=-1$) to the
difference $\Delta:=\Omega_{m0}-\Omega_{\Lambda 0}$; also, for
flat models, the constraints are similar to those from SNAP and
somewhat better than the ones from DEEP. We have also run
simulations where the ``true'' model takes into account the linear
kinematical redshift distortion and the simulated ones do not; it
then turns out that the ``true'' model is not reliably recovered
(at $2\sigma$ level), demonstrating the importance of taking this
effect explicitly into account. In contrast, for
different bias functions for the ``true'' and the simulated
models, we were able to faithfully recover $\Delta$.

We conclude by stressing that this sort of Alcock-Paczy\'nski test
is very promising and referring the reader to a more detailed
version of this work (Calv\~ao, de Mello Neto \& Waga 2001).


\begin{references}

\reference Alcock, C. \& Paczy\'nski, B. 1979, Nature 281, 358.

\reference Phillipps, S. 1994, \mnras\ 269, 1077.

\reference Calv\~ao, M. O., J. R. T. de Mello Neto \& Waga, I.
2001, astro-ph/0107029.

\reference Croom, S. M. et al.\ 2001, \mnras, in press.

\reference Goliath, M. et al.\ 2001, astro-ph/0104009.

\reference Hamilton, A. J. S. 1992, \apj\ 385, L5.

\reference Matsubara, T. \& Suto, Y. 1996, \apj\ 518, 24.

\reference Newman, J. F. \& Davis, M. 2000, \apj\ 534, L11.

\reference Perlmutter, S. et al.\ 1999, \apj\ 517, 565.

\reference Popowski, P. A., Weinberg, D. H., Ryden, B. S. \&
Osmer, P.S. 1998, \apj\ 498, 11.

\end{references}
\end{document}